\documentclass[11pt,a4paper]{article}

\usepackage{amsfonts,enumerate,subfigure,amsthm,latexsym,amssymb,amsmath,amsfonts} 
\usepackage[final]{graphicx}
\usepackage[top=1in,bottom=1in,left=1in,right=1in]{geometry}
\usepackage[table]{xcolor} 
\usepackage[colorlinks=true,linkcolor=blue,urlcolor=blue,citecolor=blue]{hyperref}
\usepackage{dcolumn}   
\usepackage{bm}        
\usepackage{multicol}

\newtheorem{theorem}{\noindent Theorem}
 
\newtheorem{lemma}[theorem]{\noindent Lemma}
\newtheorem{proposition}[theorem]{\noindent Proposition} 

\usepackage[algoruled,vlined]{algorithm2e}
\SetKwInOut{Input}{Input}
\SetKwInOut{Output}{Output} 
\usepackage[colorlinks=true,linkcolor=blue,urlcolor=blue,citecolor=blue]{hyperref}
\usepackage{multirow,multicol}

\usepackage{color}
 
\long\def\symbolfootnote[#1]#2{\begingroup%
\def\thefootnote{\fnsymbol{footnote}}\footnote[#1]{#2}\endgroup}
\def\B{{\mathcal B}} \def\E{{\mathcal E}}
 \def\RR{{\mathbb R}}  
\def\qh{{\hat q}} \def\th{{\hat t}}
\def\vol{\mbox{vol}}
\def\Line{{\tt Line}}
\def\Walk{{\tt Walk}}
\def\vol{\mbox{vol}}
\def\Min{{\it min}} 
\def\Max{{\it max}}

\title{Efficient Random-Walk Methods for \\ Approximating Polytope Volume}
\author{Ioannis\,Z.~Emiris\thanks{
Department of Informatics and Telecommunications,
National and Kapodistrian University of Athens, Greece. 
\{emiris,vfisikop\}@di.uoa.gr.}
\and Vissarion Fisikopoulos\footnotemark[1]}
\date{}

\begin{document}
\maketitle
\thispagestyle{empty}

\begin{abstract} 
We experimentally study the fundamental problem of computing the volume of a convex polytope given as an intersection of linear inequalities. 
We implement and evaluate practical randomized algorithms for accurately
approximating the polytope's volume in high dimensions (e.g.\ one hundred).  
To carry out this efficiently we experimentally correlate the effect
of parameters, such as random walk length and number of sample points, on accuracy and runtime.
Moreover, we exploit the problem's geometry by implementing an iterative rounding procedure, computing partial generations of random points and designing fast polytope boundary oracles. 
Our publicly available code is significantly faster than exact computation and more accurate
than existing approximation methods. We provide volume approximations for the Birkhoff polytopes $\B_{11},\dots, \B_{15}$, whereas exact methods have only computed that of $\B_{10}$. 

\medskip

{\bf Keywords:} volume approximation, general dimension, random walk,
polytope oracle, algorithm engineering, ray shooting 
\end{abstract}

\newpage
\setcounter{page}{1}
\section{Introduction}\label{sec:intro}

A fundamental problem in discrete and computational geometry is
to compute the volume of a convex body in general dimension or,
more particularly, of a polytope.
In the past~15 years, randomized algorithms for this problem have
witnessed a remarkable progress.
Starting with the breakthrough poly-time algorithm
of~\cite{DyerFrKa91}, subsequent results brought down
the exponent on the dimension from~27 to~4~\cite{LovaszVemp06vol}.
However, the question of an efficient implementation had remained open.  

\textbf{Notation.}
Convex bodies are typically given by a membership oracle.
A polytope $P\subseteq\RR^d$ can also be represented as the convex hull
of vertices (V-polytope) or, as is the case here, as the (bounded) intersection
$$P:= \{x\in\RR^d \ |\ Ax\leq b\}$$ of $m$ halfspaces given by
$A\in\RR^{m\times d}$, $b\in\RR^m$ (H-polytope);
$\partial P$ is its boundary, and
$O^*(\cdot)$ hides polylog factors in the argument.
The input includes approximation factor $\epsilon>0$;
$W$ denotes the most important runtime parameter, namely random walk length.

\textbf{Previous work.} 
Volume computation is $\#$-P hard for V- and for H-polytopes~\cite{DyerFr88}.
Several exact algorithms are surveyed in
\cite{FukVolume00} and implemented in {\tt VINCI}~\cite{vinci},
which however cannot handle general polytopes for dimension $d>15$.
An interesting challenge is the volume of the $n$-Birkhoff polytope,
computed only for $n\le 10$ using highly specialized software
(Sect.~\ref{sec:exper}).
Regarding deterministic approximation,
no poly-time algorithm can compute the volume with less than
exponential relative error~\cite{Elekes86}.
The algorithm of~\cite{Henk93} has error $\le d!$. 

The landmark randomized poly-time algorithm in~\cite{DyerFrKa91} 
approximates the volume of a convex body with high probability 
and arbitrarily small relative error.
The best complexity, as a function of $d$, given a membership oracle, is $O^*(d^4)$ oracle calls~\cite{LovaszVemp06vol}. 
All approaches except~\cite{LovaszVemp06vol}
define a sequence of co-centric balls,
and produce uniform point samples in their intersections with $P$
to approximate the volume of $P$.

Concerning existing software (cf Sect.~\ref{sec:further}),
\cite{CousinsV13_matlab} presented recently {\tt Matlab} code 
based on~\cite{LovaszVemp06vol} and~\cite{CousinsV13}. The latter offers a
randomized algorithm for Gaussian volume
(which has no direct reduction to or from volume)
in $O^*(d^3)$, as a function of $d$.
In \cite{LovaszD12} they implement~\cite{LovaszVemp06vol}, focusing on
variance-decreasing techniques, and
an empirical estimation of mixing time.
In~\cite{LZZ07DirectMC}, they use a straightforward acceptance-rejection
method, which is not expected to work in high dimension;
it was tested only for $d\le 4$.
An approach using thermodynamic integration~\cite{Jaekel11} offers only
experimental guarantees on runtime and accuracy.

The key ingredient of all approaches is random walks that produce an
almost uniform point sample.  Such samples is a fundamental problem
of independent interest with important applications in, e.g.,
global optimization, statistics, machine learning,
Monte Carlo (MC) integration, and non-redundant constraint identification.
Several questions of sampling combinatorial structures such as
contingency tables 
and more generally lattice points in polytopes 
may be reduced to sampling a polytope.

No simple sampling method exists unless the body
has standard shape, e.g., simplex, cube or ellipsoid.
Acceptance-rejection techniques are inefficient in high dimensions.
E.g., the number of uniform points one needs to generate in a 
bounding box before finding one in $P$ is exponential in $d$.
A 
Markov chain is the only known method, and it
may use geometric random walks such as the grid walk,
the ball walk (or variants such as the Dikin walk),
and Hit-and-run \cite{Simonovits03}. 
The Markov chain has to make a (large) number of steps, before the
generated point becomes distributed approximately uniformly
(which is the stationary limit distribution of the chain).
We focus on Hit-and-run which yields the fastest algorithms today.  

In contrast to other walks, Hit-and-run is implemented by 
computing the intersection of a line with $\partial P$.  
In general, this reduces to binary search on the line,
calling membership at every step. 
For H-polytopes, the intersection is obtained by a {\em boundary oracle};
for this, we employ ray-shooting with respect to
the $m$ facet hyperplanes (Sect.~\ref{sec:OraclSampl}).
In exact form, it is possible to avoid linear-time queries 
by using space in $o(m^{\lfloor d/2 \rfloor})$,
achieving queries in $O(\log m)$~\cite{Ramos99}.
Duality reduces oracles to (approximate) $\varepsilon$-nearest neighbor
queries, which take 
$O(dm^{(1+\varepsilon)^{-2}+o(1)})$
using $O(dm + m^{1+(1+\varepsilon)^{-2}+o(1)})$ space by locality sensitive hashing~\cite{Andoni08}.
Moreover, space-time tradeoffs from $O(1/\varepsilon^{(d-O(1))/8})$ time
and $O(1/\varepsilon^{(d-O(1))/2})$ space to $O(1)$ time and 
$O(1/\varepsilon^{(d-O(1))})$ space are available by~\cite{AryaFonsMoun12socg}.
Approximate oracles are also connected to polytope approximation.
Classic results, such as Dudley's, 
show that $O((1/\varepsilon)^{(d-1)/2})$ facets 
suffice to approximate a convex body of unit diameter
within a Hausdorff distance of $\varepsilon$.
This is optimized 
to $O(\sqrt{\vol( \partial P )}/\varepsilon^{(d-1)/2})$ 
\cite{AryaFonsMoun12socg}.
The boundary oracle is dual to finding the extreme point in
a given direction among a known pointset.
This is $\varepsilon$-approximated through $\varepsilon$-coresets
for measuring extent, in particular (directional) width, but requires
a subset of $O((1/ \varepsilon)^{(d-1)/2})$ points
\cite{AgarwalHPVa05}.  
The exponential dependence on $d$ or the linear dependence on $m$
make all aforementioned methods of little practical use.
Ray shooting has been studied in practice only in low dimensions,
e.g., in 6-dimensional V-polytopes~\cite{ZhenYaman13}. 

\textbf{Contribution.}
We implement and experimentally study efficient algorithms for approximating
the volume of polytopes. 
Point sampling, which is the bottleneck of these algorithms,
is key in achieving poly-time complexity and high accuracy.
To this end, we study variants of Hit-and-run. 
It is widely believed 
that the theoretical bound on $W$ is quite loose,
and this is confirmed by our experiments, where we set $W = O(d)$
and obtain a $<2\%$ error in up to $100$ dimensions (Sect.~\ref{sec:exper}).  

Our emphasis is to exploit the underlying geometry.
Our algorithm uses the recursive technique of co-centric balls (cf.\ Sect.~\ref{sec:vol}) introduced in~\cite{DyerFrKa91} and used in a series of papers, with the most recent to be~\cite{KannanLS97}.  
This technique forms a sequence of {\em diminishing} radii
which, unlike previous papers, allowing us to only sample {\em partial generations} of points in each intersection with $P$, instead of sampling $N$ points for each.
In fact, the algorithm starts with computing the largest interior ball by an LP.
Unlike most theoretical approaches, that use an involved rounding procedure,
we sample a set of points in $P$ and compute the minimum enclosing ellipsoid of this set, which is then linearly transformed to a ball. This procedure  is repeated until the ratio of the minimum over the maximum ellipsoid axes reaches some user-defined threshold. This {\em iterative rounding} allows us to handle skinny polytopes efficiently.

We study various oracles (Sect.~\ref{sec:OraclSampl}). 
Line search using membership requires
$O(md + \log \frac{r}{\epsilon_s})$ arithmetic operations.
This is improved to a boundary oracle in $O(md)$ by avoiding membership. 
Using Coordinate Direction Hit-and-run, we further improve 
the oracle to $O(m)$ amortized complexity.
We also exploit duality to reduce the oracle to
$\varepsilon$-nearest neighbor search:
although the asymptotic complexity is not improved,
for certain instances such as cross-polytopes in $d=16$,
kd-trees achieve a 40x speed-up. 

Our C++ code is
open-source (sourceforge) and uses the {\tt CGAL} library. A series of experiments establishes that
it handles dimensions substantially larger than existing exact approaches,
e.g., cubes and products of simplices
within an error of $2\%$ for $d\le 100$, in about 20~min.
Compared to approximate approaches, it computes significantly
more accurate results.
It computes in few hours volume estimations within an error of $2\%$ for Birkhoff polytopes $\B_{2},\dots,\B_{10}$; $\vol(\B_{10})$ has been exactly computed by specialized parallel software in a sequential time of years.   
More interestingly, it provides volume estimations for \vol$(\B_{11})$,\dots,\vol$(\B_{15})$, whose exact values are unknown, within 9 hours.  
In conclusion, we claim that the volume of general H-polytopes
in high dimensions (e.g.\ one hundred) can be efficiently and accurately
approximated on standard computers.  

\textbf{Paper organization.}
The next section discusses walks and oracles.
Sect.~\ref{sec:vol} presents the overall volume algorithm.  
Sect.~\ref{sec:exper} discusses our experiments, and we conclude
with open questions in Sect.~\ref{sec:further}.

\section{Random walks and Oracles} \label{sec:OraclSampl}
This section introduces the paradigm of Hit-and-run walks and focuses on
their implementation, with particular emphasis on exploiting the geometry
of H-polytopes. The methods presented here are analysed experimentally in Sect.~\ref{sec:exper}.

\textbf{Hit-and-run random walks.} 
The main method to randomly sample a polytope is
by (geometric) random walks.
We shall focus on variants of Hit-and-run, which generate a uniform
distribution of points~\cite{smith1984montecarlo}. 
Assume we possess procedure \Line($p$), which returns line $\ell$ through
point $p\in P\subseteq\RR^d$; $\ell$ will be specified below.
The main procedure of Hit-and-run is \Walk($p,P,W$), which reads in point
$p\in P$ and repeats $W$ times:
(i) run \Line($p$), 
(ii) move $p$ to a random point uniformly distributed on $P\cap \ell$.
We shall consider two variants of Hit-and-run.

In {\it Random Directions Hit-and-run} (RDHR),
\Line($p$) returns $\ell$ defined by a random vector
uniformly distributed on the unit sphere centered at $p$.
The vector coordinates are drawn from the standard normal distribution.
RDHR generates a uniformly distributed point in  
\begin{eqnarray}\label{Ewalklength}
O^*(d^2 r^2),\, \mbox{ or }\, O^*(d^3 r^2)\, \mbox{ oracle calls},\\ 
\mbox{ with hidden constants }\, 10^{30},\, \mbox{ or }\, 10^{11} \mbox{ respectively, }\nonumber
\end{eqnarray}
starting at an arbitrary, or at
a uniformly distributed point (also known as warm start), respectively, 
where $r$ is the ratio of the radius of the smallest enclosing 
ball over that of the largest enclosed ball in $P$~\cite{LovaszV06corner}.

In {\it Coordinate Directions Hit-and-run} (CDHR),
\Line($p$) returns $\ell$ defined by a random vector
uniformly distributed on the set $\{e_1,\dots,e_d\}$,
where $e_i=(0,\dots,0,1,0,\dots,0),\ i=1,\dots,d.$
This is a continuous variant of the Grid walk.  
As far as the authors know, the mixing time has not been analyzed.
We offer experimental evidence that CDHR is faster than RDHR and
sufficiently accurate.
An intermediate variant is Artificially Centering
Hit-and-run~\cite{Smith_AHNR98}, where first a set $S$ of sample points
is generated as with RDHR, then \Line($p$) returns $\ell$ through $p$ and a randomly
selected point from $S$.
This however is not a Markov chain, unlike CDHR and RDHR.

Procedure \Walk($p$,\ $P$,\ $W$) requires at every step an access to a {\it boundary oracle} which computes the intersection of line $\ell$ with $\partial P$.
In the sequel we discuss various implementations of this oracle. 

\vspace{.5em}
\textbf{Boundary oracle by membership.}
For general convex bodies, a boundary oracle can be implemented using a {\em membership oracle} which, given vector $y\in \RR^d$, decides whether $y\in P$.
The intersection of $\ell$ with $\partial P$ is computed
by binary search on the segment defined by any point on $\ell$ lying in
the body and the intersection of $\ell$ with a bounding ball.
Each step calls membership to test whether the current point is internal,
and stops when some accuracy $\epsilon_s$ is certified.
Checking the point against a hyperplane takes $O(d)$ operations,
thus obtaining the intersection of $\ell$ with the hyperplane.
We store this intersection so that subsequent tests against
this hyperplane take $O(1)$.
The total complexity is $O(md+\log \frac{r}{\epsilon_s})$ arithmetic operations,
where $r$ is the ball radius. 

\textbf{Boundary oracle by facet intersection.}
Given an H-polytope $P$ the direct method to compute the intersection of line $\ell$ with $\partial P$ is to examine all $m$ hyperplanes.
Let us consider \Walk($p_0,P,W$) and
line $\ell = \{ x \in\RR^d : x =\lambda v+ p_0 \}$,
where $p_0\in\RR^d$ lies on $\ell$, and $v$ is the 
direction of $\ell$.
We compute the intersection of $\ell$ with
the $i$-th hyperplane ${a_ix}={b_i}$, $a_i\in\RR^d,b_i\in\RR$, namely
$
{p_i} := { p_0} + \frac{b_{i}-{a_i} { p_0}}{{a_i v}}{v},
\, i\in\{1,\dots,m\} .
$
We seek points $p^+,p^-$ at which $\ell$ intersects $\partial P$,
namely ${p^+v} = \min_{1\le i\le m}\{{p_iv} \ |\ {p_iv} \ge 0\}$  and 
${p^-v} = \max_{1\le i\le m}\{{p_iv} \ |\ {p_iv} \le 0\}.$
This is computed in $O(md)$ arithmetic operations.  
In practice, only the $\lambda^{\pm}$ are computed, where
${p^{\pm}}={p_0}+\lambda^{\pm} {v}$.

In the context of the volume algorithm (Sect.\ref{sec:vol}), 
the intersection points of $\ell$ with $\partial P$ are compared to the intersections of $\ell$ with the current sphere.
Assuming the sphere is centered at the origin with radius $R$,
its intersections with $\ell$
are ${p}={p_0}+\lambda {v}$ such that 
$\lambda^2 + 2\lambda {p_0}{v} + |{p_0}|^2 - R^2 = 0$. 
If $\lambda^+,\lambda^-$ give a negative sign when substituted to
the aforementioned equation 
then ${p^+,p^-}$ are the
endpoints of the segment of $\ell$ lying in the intersection of
$P$ and the current ball.
Otherwise, we have to compute one or two roots of
the aforementioned equation 
since the segment has one or two endpoints on the sphere.

However, in CDHR, where $\ell$ and $v$ are vertical,  
after the computation of the first pair ${p^+},{p^-}$, all
other pairs can be computed in $O(m)$ arithmetic operations.
This is because two sequential points produced by the walk
differ only in one coordinate.
Let $j,k$ be the walk coordinate of the previous and the current
step respectively.
Then, assuming $P=\{x\in\RR^d: Ax\le b\}$, where $A\in\RR^{m\times d}$,
$
\lambda^{\pm} = \max\{ \lambda\; |\; A ( p_0\pm\lambda v ) \le b \}.
$
This becomes $\pm \lambda Av = \pm \lambda A_{j} \le -Ap_0 + b $,
where $A_{j}$ is the $j$-th column of $A$.
The two maximizations are solved in $O(md)$ ops.
Let vector $t=-Ap_0 + b\in\RR^m$.
At the next step, given point $p_0'=p_0+ce_{j}$, where $e_{j}$
is the $j$-th standard basis vector, we perform two maximizations
$\lambda : \pm \lambda A_{k} \le t - cA_{j}$ in $O(m)$.

\textbf{Boundary oracle by duality.}\label{subsec:duality}
Duality reduces the problem to nearest neighbor (NN) search and its variants. 
Given a pointset $B\subseteq\RR^d$ and query point $q$, NN search returns a point $p\in B$ 
s.t.\ $dist(q, p) \leq dist(q, p' )$ for all $p'\in B$, where $dist(q, p )$
is the Euclidean distance between points $q, p$.
Let us consider, w.l.o.g., boundary intersection for line $\ell$
parallel to the $x_d$-axis:
$ \ell = \{ x : x=\lambda v+ p,\, \lambda\ge 0 \}, \, v=(0,\dots,0,-1).$
It reduces to two ray-shooting questions; it suffices to describe one,
namely with the upward 
vertical ray, defined by $\lambda\le 0$.  
We seek the first facet hyperplane hit which, equivalently, has the
maximum negative signed vertical distance from $p$ to any hyperplane $H$
of the upper hull, for fixed $v$.
This distance is denoted by sv$(p, H)$.
Let us consider the standard (aka functional) duality transform
between points $p$ and non-vertical hyperplanes $H$:
\begin{small}
\begin{eqnarray*}
p =(p_1,\dots,p_d)   \mapsto
	p^*:x_d = p_1x_1 + \cdots + p_{d-1}x_{d-1} - p_d, \\
H: x_d = c_1x_1 + \cdots + c_{d-1}x_{d-1} + c_0 
	 \mapsto  H^* =(c_1,\dots, c_{d-1}, -c_0).
\end{eqnarray*}
\end{small}
This transformation is self-dual, preserves point-hyperplane incidences,
and negates vertical
distance, hence sv$^*(p^*,H^*)=-\mbox{sv}(p,H)$, where sv$^*(\cdot,\cdot)$
is the signed vertical distance from hyperplane $p^*$ to point $H^*$
in dual space.
Hence, our problem is equivalent to
minimizing sv$^*(p^*,H^*) \ge 0$.
Equivalently, we seek point $H^*$ minimizing absolute vertical distance 
to hyperplane $p^*$ on its side of 
positive distances.
In dual space, consider 
\begin{align}\label{Eduals}
\mbox{point } t =(t_1,\dots,t_d), \mbox{ and hyperplane }\nonumber\\
p^* = q: x_d=q_1x_1+\cdots + q_{d-1}x_{d-1}+ q_0 :
\end{align}
\begin{align*}
\mbox{sv}^*(q, t) &= t_d- (q_1t_1+\cdots + q_{d-1}t_{d-1} +q_0 )\nonumber\\
 &= - (q_0,q_1,\dots ,q_{d-1} , -1) \cdot (1, t_1, \dots, t_{d-1}, t_d) ,
\end{align*}
where the latter operation is inner product in Euclidean space $\RR^{d+1}$
of ``lifted" datapoint $t' =(1, t_1, \dots, t_{d-1}, t_d)$ with
``lifted" query point $q' =(q_0,q_1,\dots ,q_{d-1} , -1)$.
Let 
$$
q'' =(q',0), \; t'' =(t',\sqrt{M-\|t'\|_2^2}), \, \mbox{ for }
M\ge 
\max_t \{ 1 + \|t\|_2^2 \} ,
$$
following an idea of \cite{BasriHZM11}. By the cosine rule, 
$$
\mbox{dist}_{d+2}^2 (q'',t'') = 
\| q' \|_2^2 + M + 2 \mbox{sv}^*(q,t) , 
$$
where dist$_{d+2}(\cdot,\cdot)$ stands for Euclidean distance in $\RR^{d+2}$.
Since the $t''$ lie on hyperplane $x_1=1$, optimizing
$\mbox{dist}_{d+2}(q'',t'')$ over a set of points $t''$ is equivalent to optimizing
$ \mbox{dist}_{d+1}(\qh,\th)$, $\qh=(q_1,\dots,q_{d-1},-1,0),
\mbox{ over points } \th=(t,\sqrt{M-1-\|t\|_2^2}).$ 
Hence, point $t$ 
minimizing sv$^*(q,t) \ge 0$ corresponds to $\th$
minimizing dist$_{d+1}^2 (\qh,\th)$. 
Thus the problem is reduced to (exact) nearest neighbor in $\RR^{d+1}$.
Ray shooting to the lower hull with same $v$ reduces to 
farthest neighbor. 
Unfortunately, an approximate solution to these problems incurs
an additive error to the corresponding original problem.  

Alternatively, we shall consider hyperplane queries.
Let us concentrate on hyperplanes supporting facets on the lower 
hull of $P$.  Their dual points lie in convex position.
Given that point $p$ is interior in $P$, the dual points
of the lower hull facets lie on the upper halfspace of $p^*$.
In dual space, consider point $t$ 
and hyperplane $q$ 
as in expression~(\ref{Eduals}).
Let sd$^*(q,t)$ be the signed Euclidean distance from $q$ to $t$, i.e.\
the minimum Euclidean distance of any point on $q$ to $t$.
Then
$
\mbox{sv}^*(q, t) = \mbox{sd}^*(q,t)\, / \, \| (q_1,\dots ,q_{d-1} , 1) \|_2 ,
$
where the normal is $(q_1,\dots ,q_{d-1} , 1)$.
Our question, therefore, becomes equivalent to 
minimizing sd$^*(q,t)$ over all datapoints $t\in\RR^d$ for which
sd$^*(q,t)\ge 0$; i.e., we seek the NN above $q$.
Starting with facets on the upper hull, the problem becomes that of
maximizing sd$^*(q,t)\le 0$, i.e.\ finding the NN below $q$.

The above approaches motivate us to use NN software for exact 
point and hyperplane queries (Sect.~\ref{sec:exper}).

\section{The volume algorithm}\label{sec:vol}

This section details our poly-time methods for approximating the volume of $P$.
Algorithms in this family are the current state-of-the-art with
respect to asymptotic complexity bounds.
Moreover, they can achieve any approximation ratio given by the user,
i.e., they form a fully polynomial randomized approximation scheme (FPRAS).
Given polytope $P\subseteq\RR^d$, they execute 
{\it sandwiching} and {\it Multiphase Monte Carlo} (MMC)~\cite{Simonovits03}. 

We consider that $P$ is a full-dimensional $H$-polytope. However, we can also consider $P$ to be lower dimensional and be given in form 
$\{x\in\RR^{d} \, |\, Ax=b,\, x\geq 0\}$, where $A\in\RR^{m\times d}$, $x\in\RR^{d}$,  $b\in\RR^{m}$, $A'\in\RR^{m\times m-d+1}$,  $x'\in\RR^{m-d+1}$. Using {\it Gauss-Jordan elimination} the linear system $Ax=b$ can be transformed to its unique {\it reduced row echelon form} $[I|A']x=b'$, where $I$ is the identity matrix.  Then $P$ can be written as $\{x'\in\RR^{m-d+1} \ |\ A'x'\geq b', x'\geq 0\}$, i.e.\ a full-dimensional $H$-polytope in $\RR^{m-d+1}$. 

\textbf{Rounding and sandwiching.}
This stage involves first rounding $P$ to reach a
near isotropic position, second sandwiching, i.e.\ to compute ball $B$
and scalar $\rho$ such that $B\subseteq P\subseteq\rho B$.
There is an abundance of methods in literature for rounding and sandwiching (cf.~\cite{Simonovits03} and references therein). However, here we develop a simple, efficient method that succeed significantly accurate results in practice (cf.\ Sect.~\ref{sec:exper} and Table~\ref{table:rounding}). The method doesn't compute a ball that covers $P$ but a ball $B'$ such that $B'\cap P$ contains almost all the volume of $P$.

For rounding, we sample a set $S$ of $O(n)$ random points in $P$.
Then we approximate the minimum volume ellipsoid $\E$ that covers $S$,
and satisfies the inclusions
$\frac{1}{(1+\varepsilon)d}\E\subseteq \mbox{conv}(S)\subseteq \E$,
in time $O(nd^2(\varepsilon^{-1}+\ln d+\ln\ln n))$ \cite{Khachiyan96}.
Let us write
\begin{align}\label{eq:ellipsoid}
 \E&=\{x\in\RR^d\,|\,(x-c_{\E})^T\,E\,(x-c_{\E})\leq 1\}\nonumber\\
   &=\{x\in\RR^d\,|\,L^T(x-c_{\E})\leq 1\},
\end{align}
where $E\subseteq\RR^{d\times d}$ is a positive semi-definite (p.s.d.) matrix and
$L^TL$ its Cholesky decomposition.
By substituting $x=(L^T)^{-1}y+c_{\E}$ we map the ellipsoid to the ball 
$\{y\in\RR^d\,|\,y^Ty\leq 1\}$. Applying this transformation to $P$ we have $P'=\{y\in\RR^d\,|A(L^T)^{-1}\leq b-Ac_\E\}$ which is the rounded polytope,
where \vol$(P)= \det(L^T)^{-1} \vol(P')$. 
We iterate this procedure until the ratio of the minimum over the maximum ellipsoid axes reaches some user defined threshold.

For sandwiching $P$ we first compute the
{\it Chebychev ball} $B(c,r)$ of $P$, i.e.\ the largest inscribed ball in $P$.
It suffices to solve the LP:
$\{
\text{maximize } R , \; 
\text{ subject to: } A_i x+R \|A_i\|_2\leq b_i,\, i=1,\dots,m, \; R\geq 0\},
$
where $A_i$ is the $i$-th row of $A$, and
the optimal values of $R$ and $x\in\RR^d$ yield, respectively,
the radius $r$ and the center $c$ of the Chebychev ball.

Then we may compute a uniform random point in $B(c,r)$ and use it as a start
to perform a random walk in $P$, eventually generating $N$ random points.
Now, compute the largest distance between each of the $N$ points and $c$;
this defines a (approximate) bounding ball.
Finally, define the sequence of balls  
$
B(c,2^{i/d}),\, i= \alpha, \alpha+1 , \dots, \beta,
$
where $\alpha=\lfloor d\log r \rfloor$ and $\beta=\lceil d\log \rho\rceil$. 

\textbf{Multiphase Monte Carlo (MMC).}
MMC constructs a sequence of bodies 
$P_i:=P\cap B(c,2^{i/d}),\; i=\alpha, \alpha+1 , \dots, \beta,$
where $P_{\alpha}=B(c,2^{\alpha/d})\subseteq B(c,r)$ and $P_{\beta}$
(almost) contains $P$.
Then it approximates vol$(P)$ by the telescopic product
$$
\vol(P_{\alpha})\prod_{i=\alpha+1}^{\beta} \frac{\vol(P_i)}{\vol(P_{i-1})},
\,\, 
\mbox{where\, \vol}(P_{\alpha})= \frac{2\pi^{d/2}(2^{\lfloor\log r \rfloor})^d}{d\,\Gamma(d/2)}.
$$
This reduces to estimating the ratios
${\vol(P_i)}/{\vol(P_{i-1})}$, which is achieved by 
generating $N$ uniformly distributed points in ${P_i}$
and by counting how many of them fall in ${P_{i-1}}$.

For point generation we use random walks as in Sect.~\ref{sec:OraclSampl}. 
We set the walk length $W=\lfloor 10+d/10\rfloor=O(d)$, which is of the same
order as in \cite{LovaszD12} but significantly lower than theoretical bounds.
This choice is corroborated experimentally (Sect.~\ref{sec:exper}). 

Unlike typical approaches, which generate points in $P_i$
for $i=\alpha,\alpha+1,\dots,\beta$, here we proceed inversely.
First, let us describe initialization.
We generate an (almost) uniformly distributed random point $p\in P_\alpha$,
which is easy since $P_\alpha=B(c,2^{\alpha/d})\subseteq B(c,r)$.
Then we use $p$ to start a random walk in $P_\alpha,P_{\alpha+1},P_{\alpha+2}$
and so on,
until we obtain a uniformly distributed point in $P_{\beta}$. 
We perform $N$ random walks starting from this point to generate $N$
(almost) uniformly distributed points in $P_{\beta}$ and then
count how many of them fall into $P_{\beta-1}$.
This yields an estimate of $\vol(P_{\beta})/\vol(P_{\beta-1})$.
Next we keep the points that lie in $P_{\beta-1}$,
and use them to start walks so as to gather a total of $N$ (almost) 
uniformly distributed points in $P_{\beta-1}$.
We repeat until we compute the last ratio ${\vol(P_{\alpha+1})}/{\vol(P_{\alpha})}$. 

The implementation is based on
a data structure $S$ that stores the random points.
In step $i>\alpha$, we wish to compute
${\mbox{vol}(P_{\beta-i})}/{\mbox{vol}(P_{\beta-i-1})}$ and $S$ contains $N$ random points in $P_{\beta-i+1}$ from the previous step.
The computation in this step consists in removing from $S$ the points not in
$P_{\beta-i}$, then sampling $N-size(S)$ new points in $P_{\beta-i}$
and, finally, counting how many lie in $P_{\beta-i-1}$.
Testing whether such a point lies in some $P_{i}$
reduces to testing whether $p\in B(2^{i/d})$ because $p\in P$.

One main advantage of our method is that it creates partial generations
of random points for every new body $P_i$, as opposed to having
always to generate $N$ points.
This has a significant effect on runtime since it reduces it by a
constant raised to $\beta$.
Partial generations of points have been used
in convex optimization~\cite{BertsimasVempala04}.
 
\def\Plarge{P_{\mbox{\footnotesize large}}}{}
\def\Psmall{P_{\mbox{\footnotesize small}}}{} 
\begin{algorithm}[t!] \caption{\label{Alg:vol} VolEsti ($P,\epsilon,t_r$)}
 \BlankLine
  \Input{H-polytope $P$, objective approximation $\epsilon$, rounding threshold $t_r$} 
  \Output{approximation of \vol$(P)$}
  \BlankLine
  $N\leftarrow 400\epsilon^{-2}d\log d$;\quad 
  $W\leftarrow \lfloor 10+d/10\rfloor$\;  
  \BlankLine\tcp{rounding and sandwiching}  
  compute the Chebychev ball $B(c,r)$\; 
  generate a random point $p$ in $B(c,r)$\;
  \Repeat{$\E max/ \E min<t_r$}{    
    $S\leftarrow \emptyset$\;  
    \For{$i=1$ to $N$}{
      $p\leftarrow$ \Walk($p,P,W$)\;
      add $p$ in $S$\;
    }
    compute min encl.\ ellipsoid $\E$ of $S$, with p.s.d.\ $E$\;
    set as $\E min, \E max$ the min and max $\E$  axes\;
    compute the Cholesky decomposition $L^TL$ of $E$\;
    transform $P$ and $p$ w.r.t.\ $L$\;
  }  
  set $\rho$ the largest distance from $c$ to any point in $S$\;
  \BlankLine  
  \tcp{MMC}  
  set $\alpha\leftarrow \lfloor\log r \rfloor$;\quad
  $\beta\leftarrow  \lceil\log \rho\rceil$\;
  $P_i\leftarrow P\cap B(c,2^{i/d})$ for $i=\alpha, \alpha+1 , \dots, \beta$\;
  $\vol(P_{\alpha})\leftarrow
	2\pi^{d/2}(2^{\lfloor\log r \rfloor})^d/d \,\Gamma(d/2)$\;    
  $i\leftarrow\beta$\; 
  \While{$i>\alpha$}{
    $\Plarge\leftarrow P_i$;\quad
    $i\leftarrow i-1$;\quad
    $\Psmall\leftarrow P_i$\;    
    $count\_prev\leftarrow size(S)$;\quad 
    remove from $S$ the points not in $\Psmall$;\quad
    $count\leftarrow size(S)$\; 
    Set $p$ to be an arbitrary point from $S$\;   
    \For{$j=1$ to $N-count\_prev$}{
      $p\leftarrow$ \Walk($p,\Plarge,W$)\;
      \If{$p\in B(c,2^{i/d})$}{
        $count\leftarrow count +1$\;
        add $p$ in $S$\;
      }
    }
    $vol\leftarrow vol \cdot (N/count)$\;
  }  
  \Return $\vol/\det(L^T)$ \;
  \BlankLine
\end{algorithm} 
 
We use {\it threads}, also in~\cite{LovaszD12}, to ensure independence
of the points. A thread is a sequence of points each generated from
the previous point in the sequence by a random walk.
The first point in the sequence is uniformly distributed in the ball
inscribed in $P$.
Alg.~\ref{Alg:vol} describes our algorithm using a single thread. 

\textbf{Complexity.}
The first $O^*(d^5)$ algorithm was in~\cite{KannanLS97}, using
a sequence of subsets defined as the intersection of
the given body with a ball. 
It uses isotropic sandwiching to bound the number of balls by $O^*(d)$,
it samples $N= 400\epsilon^{-2}d\log d=O^*(d)$ points per ball, 
and follows a ball walk to generate each point in $O^*(d^3)$ oracle calls. 
Interestingly, both sandwiching and MMC each require $O^*(d^5)$ oracle calls.
Later 
the same complexity was obtained by
Hit-and-run under the assumption the convex body is well sandwiched.

\begin{proposition}{\em\cite{KannanLS97}}\label{prop:volume}
Assuming $B(0,1) \subseteq P \subseteq  B(0,\rho)$, the volume algorithm of~\cite{KannanLS97} 
returns an estimation of \vol$(P)$, which lies between 
$(1-\epsilon)\vol(P)$ and $(1+\epsilon)\vol(P)$,
with probability $\ge 3/4$, by 
$$
O\left(\frac{d^4\rho^2}{\epsilon^2}\ln d
	\ln \rho\ \ln^2\frac{d}{\epsilon} \right)=O^*(d^4\rho^2)
$$ 
oracle calls with probability $\ge 9/10$, where we have assumed
$\epsilon$ is fixed.
Sandwiching yields
$\rho=\sqrt{d/\log(1/\epsilon)}$, implying a total of $O^*(d^5)$ calls.
\end{proposition}

In~\cite{LovaszVemp06vol}, they construct 
a sequence of log-concave functions and estimate ratios of integrals, instead of ratios of balls, using simulated annealing.
The complexity reduces to $O^{*}(d^{4})$
by decreasing both number of phases 
and number of samples per phase to $O^*(\sqrt{d})$. Using Hit-and-run, $O^{*}(d^{3})$ still bounds the time to sample each point.
Moreover, they improve isoperimetric sandwiching to $O^{*}(d^{4})$.
 
The following Lemma states the runtime of Alg.~\ref{Alg:vol}, which is in fact a variant of the algorithm analysed in~\cite{KannanLS97} (see also Prop.~\ref{prop:volume}). Although there is no theoretical bound on the approximation error of Alg.~\ref{Alg:vol}, our experimental analysis in Sect.~\ref{sec:exper} shows that in practice the achieved error is always  better than the one proved in~Prop.~\ref{prop:volume}.

\begin{lemma}
Given $H$-polytope $P$, Alg.~\ref{Alg:vol}
performs $k$ phases of rounding in $O^*(d^3mk)$, and approximates $\vol(P)$
in $O(md^3\log d\log (\rho/r))$ arithmetic operations,
assuming $\epsilon>0$ is fixed,
where $r$ and $\rho$ denote the radii of the largest inscribed ball and
of the co-centric ball covering $P$.
\end{lemma}
\begin{proof}
Our approach generates $d\log (\rho/r)$ balls and uses Hit-and-run.
Assuming $P$ contains the unit ball, an upper bound on $\rho/r$
is diameter $\delta$. 
In each ball intersected with $P$, we generate
$\le N= 400\epsilon^{-2}d\log d$ random points.
Each point is computed after $W=O(d)$ steps of CDHR.

The boundary oracle of CDHR is implemented in Sect.~\ref{sec:OraclSampl}.
In particular, $k$ CDHR steps require $O(dm+(k-1)m+kd)$ arithmetic operations. 
It holds $d=O(m)$ and $k=\Omega(d)$.
Thus, the amortized complexity of a CDHR step is $O(m)$.
Overall, the algorithm needs
$O(\epsilon^{-2}md^3\log d\log(\rho/r))$ operations.

Each rounding iteration decreases $\delta$ and runs in
$O(nd^2(\varepsilon^{-1}+\ln d+\ln\ln(n)))$, where $n$ stands for the number
of sampled points, and $\varepsilon$ is the approximation of
the minimum volume ellipsoid of Eq.~(\ref{eq:ellipsoid}).
We generate $n=O(d)$ points, each in $O(m)$ arithmetic operations.
Hence, rounding runs in $O^*(d^3mk)$, where $\varepsilon$ is fixed.
Moreover, $k$ is typically constant since $k=1$ is enough to handle,
e.g., polytopes with $\rho/r=100$ in dimension up to $20$. 
\end{proof}

\begin{figure}[t]
	\centering
	\includegraphics[width=.6\textwidth]{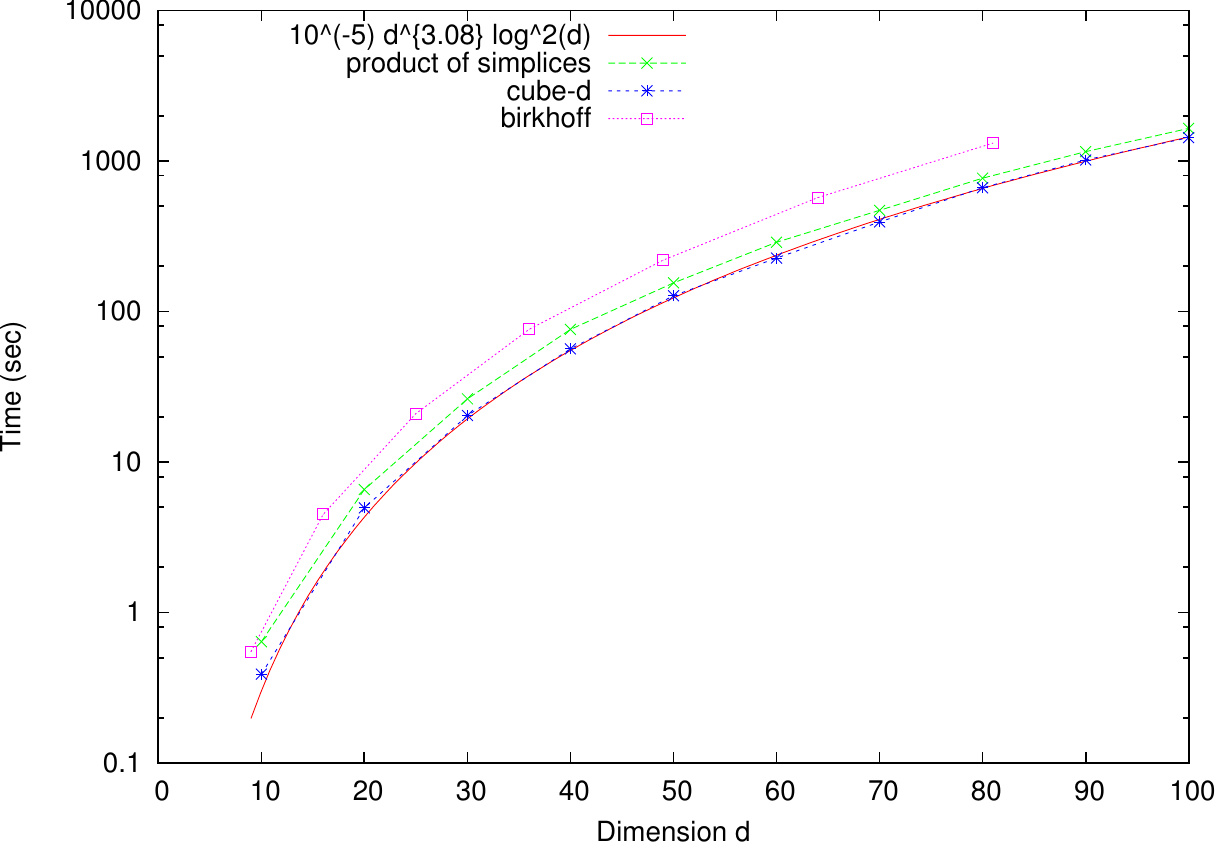}
	\caption{Experiments with $\B_5$ on the effect of $W$ and $\epsilon$ (or $N$) on
accuracy, measured by (min, max)/$\mu$ (crosses), and runtime,
measured by levels of $N\cdot W=c$, for $c=10^5,\dots, 2.5\cdot 10^6$.}\label{fig:runtime_dim}	
\end{figure}

Let us check this bound with the
experimental data for cubes, products of simplices, and Birkhoff polytopes,
with $d\le 100$ and $\epsilon=1$, where
$m=2d$, $d+2$ and $d+1+2\sqrt{d}$, respectively, for the 3 classes, 
and for cubes $\log(\rho/r) \leq \log(\sqrt{d})=O(\log d)$.  
Fig.~\ref{fig:runtime_dim} shows that the~3 classes behave similarly.
Performing a fit of $ad^b\log^2 d$, runtime follows $10^{-5}d^{3.08}\log^2 d$
which shows a smaller dependence on $d$ than our bounds,
at this range of experiments.

\section{Experiments}\label{sec:exper}


\begin{table}[t!]\centering\tiny
\begin{tabular*}{\linewidth}{@{\extracolsep{\fill}}r@{\quad}r@{\quad}r@{\quad}r
@{\quad}r@{\quad}r@{\quad}r@{\quad}r@{~}r@{\quad}r@{\quad}r}
$P$ & $d$ & $m$  & \vol$(P)$ & $N$ & $\mu$ & [min, max]  & std-dev &  
\multirow{2}{*}{$\frac{\vol(P)-\mu}{\vol(P)}$} & {\tt VolEsti} & {\tt Exact}\\
& & & & & & & & & (sec) & (sec)\\\hline
cube-10 & 10 & 20 & 1.024E+003 & 9210 & 1.027E+003 & [0.950E+003,1.107E+003] & 3.16E+001 & 0.0030 & 0.42 & 0.01\\
cube-15 & 15 & 30 & 3.277E+004 & 16248 & 3.24E+004 & [3.037E+004,3.436E+004] & 9.41E+002 & 0.0088 & 1.44 & 0.40\\
cube-20 & 20 & 40 & 1.048E+006 & 23965 & 1.046E+006 & [0.974E+006,1.116E+006] & 3.15E+004 & 0.0028 & 4.62 & swap\\
cube-50 & 50 & 100 & 1.126E+015 & 78240 & 1.125E+015 & [1.003E+015,1.253E+015] & 4.39E+013 & 0.0007 & 117.51 & swap\\
cube-100 & 100 & 200 & 1.268E+030 & 184206 & 1.278E+030 & [1.165E+030,1.402E+030] & 4.82E+028 & 0.0081 & 1285.08 & swap\\
$\Delta$-10 & 10 & 11 & 2.756E-007 & 9210 & 2.76E-007 & [2.50E-007,3.08E-007] & 1.08E-008 & 0.0021 & 0.56 & 0.01\\
$\Delta$-50 & 60 & 61 & 1.202E-082 & 98264 & 1.21E-082 & [1.07E-082,1.38E-082] & 6.44E-084 & 0.0068 & 183.12 & 0.01\\
$\Delta$-100 & 100 & 101 & 1.072E-158 & 184206 & 1.07E-158 & [9.95E-159,1.21E-158] & 4.24E-160 & 0.0032 & 907.52 & 0.02\\
$\Delta$-20-20 & 40 & 42 & 1.689E-037 & 59022 & 1.70E-037 & [1.54E-037,1.87E-037] & 7.33E-039 & 0.0088 & 53.13 & 0.01\\
$\Delta$-40-40 & 80 & 82 & 1.502E-096 & 140224 & 1.50E-096 & [1.32E-096,1.70E-096] & 7.70E-098 & 0.0015 & 452.05 & 0.01\\
$\Delta$-50-50 & 100 & 102 & 1.081E-129 & 184206 & 1.10E-129 & [1.01E-129,1.19E-129] & 4.65E-131 & 0.0154 & 919.01 & 0.02\\
cross-10 & 10 & 1024 & 2.822E-004 & 9210 & 2.821E-004 & [2.693E+004,2.944E+004] & 5.15E-006 & 0.0003 & 1.58 & 388.50\\
cross-11 & 11 & 2048 & 5.131E-005 & 10550 & 5.126E-005 & [4.888E-005,5.437E-005] & 1.15E-006 & 0.0010 & 5.19 & 6141.40\\
cross-12 & 12 & 4096 & 8.551E-006 & 11927 & 8.557E-006 & [8.130E-006,9.020E-006] & 1.69E-007 & 0.0007 & 12.21 & ---\\
cross-15 & 15 & 32768 & 2.506E-008	& 16248	& 2.505E-008 & [2.332E-008,2.622E-008] & 5.15E-010 & 0.0004 & 541.22 & ---\\
cross-18 & 18 & 262144 & 4.09E-011 & 20810 & 4.027E-011	&[3.97E-011,4.08E-011] & 5.58E-013	& 0.0165 & 5791.06 &  ---\\
rh-8-25 & 8 & 25 & 7.859E+002 & 6654 & 7.826E+002 & [7.47E+002,8.15E+002] & 1.93E+001 & 0.0042 & 0.30 & 1.14\\
rh-8-30 & 8 & 30 & 2.473E+002 & 6654 & 2.449E+002 & [2.28E+002,2.68E+002] & 1.06E+001 & 0.0099 & 0.27 & 5.56\\
rh-10-25 & 10 & 25 & 5.729E+003 & 9210 & 5.806E+003 & [5.55E+003,6.06E+003] & 1.85E+002 & 0.0134 & 0.66 & 6.88\\
rh-10-30 & 10 & 30 & 2.015E+003 & 9210 & 2.042E+003 & [1.96E+003,2.21E+003] & 7.06E+001 & 0.0132 & 0.67 & swap\\
rv-8-10 & 8 & 24 & 1.409E+019 & 6654 & 1.418E+019 & [1.339E+019,1.497E+019] & 5.24E+017 & 0.0107 & 0.37 & 0.01\\
rv-8-11 & 8 & 54 & 3.047E+018 & 6654 & 3.056E+018 & [2.562E+018,3.741E+018] & 3.98E+017 & 0.0028 & 0.76 & 0.54\\
rv-8-12 & 8 & 94 & 4.385E+019 & 6654 & 4.426E+019 & [4.105E+019,4.632E+019] & 2.07E+018 & 0.0093 & 0.59 & 261.37\\
rv-8-20 & 8 & 1191 & 2.691E+021 & 6654 & 2.724E+021 & [2.517E+021,2.871E+021] & 1.05E+020 & 0.0123 & 3.69 & swap\\
rv-8-30 & 8 & 4482 & 7.350E+021 & 6654 & 7.402E+021 & [7.126E+021,7.997E+021] & 2.19E+020 & 0.0072 & 12.73 & swap\\
rv-10-12 & 10 & 35 & 2.136E+022 & 9210 & 2.155E+022 & [1.952E+022,2.430E+022] & 1.53E+021 & 0.0093 & 1.00 & 0.01\\
rv-10-13 & 10 & 89 & 1.632E+023 & 9210 & 1.618E+023 & [1.514E+023,1.714E+023] & 6.23E+021 & 0.0088 & 1.24 & 59.50\\
rv-10-14 & 10 & 177 & 2.931E+023 & 9210 & 2.962E+023 & [2.729E+023,3.195E+023] & 1.71E+022 & 0.0135 & 2.08 & swap\\
cc-8-10 & 8 & 70 & 1.568E+005 & 26616 & 1.589E+005 & [1.52E+005,1.64E+005] & 3.50E+003 & 0.0138 & 1.95 & 0.05\\
cc-8-11 & 8 & 88 & 1.391E+006 & 26616 & 1.387E+006 & [1.35E+006,1.43E+006] & 2.65E+004 & 0.0034 & 2.10 & 0.08\\
Fm-4 & 6 & 7 & 8.640E+001 & 4300 & 8.593E+001 & [7.13E+001,1.12E+002] & 8.38E+000 & 0.0055 & 0.19 & 0.01\\
Fm-5 & 10 & 25 & 7.110E+003 & 9210 & 7.116E+003 & [6.35E+003,8.10E+003] & 3.01E+002 & 0.0009 & 0.69 & 0.02\\
Fm-6 & 15 & 59 & 2.861E+005 & 16248 & 2.850E+005 & [2.42E+005,3.22E+005] & 1.55E+004 & 0.0038 & 3.24 & swap\\
ccp-5 & 10 & 56 & 2.312E+000 & 9210 & 2.326E+000 & [2.16E+000,2.52E+000] & 7.43E-002 & 0.0064 & 0.49 & 38.00\\
ccp-6 & 15 & 368 & 1.346E+000 & 16248 & 1.346E+000 & [1.26E+000,1.45E+000] & 3.81E-002 & 0.0002 & 6.14 & swap\\
$\B_{8}$ & 49 & 64 & 4.42E-023 & 76279 & 4.46E-023 & [4.05E-023,\,7.32E-024] & 1.93E+004 & 0.0092 & 192.97 & 1920.00\\ 
$\B_{9}$ & 64 & 81 &  2.60E-033 & 106467 & 2.58E-033 & [2.23E-033,\,3.07E-033] & 2.13E-034 & 0.0069 & 499.56 & 8 days\\
$\B_{10}$ & 81 & 100 & 8.78E-046 & 142380 & 8.92E-046 & [7.97E-046,\,9.96E-046] & 4.99E-047 & 0.0152 & 1034.74 & 6160 days\\
$\B_{11}$ & 100 & 121 & ??? & 184206 & 1.40E-060 & [1.06E-060,\,1.67E-060] & 1.10E-061 & ??? & 2398.17 & ---\\
$\B_{12}$ & 121 & 144 & ??? & 232116 &  7.85E-078 & [6.50E-078,\,9.31E-078] & 5.69E-079 & ??? & 4946.42 & ---\\
$\B_{13}$ & 144 & 169 & ??? & 286261 &  1.33E-097 & [1.13E-097,\,1.62E-097] & 1.09E-098 & ??? & 9802.73 & ---\\
$\B_{14}$ & 169 & 196 & ??? & 346781 &  5.96E-120 & [5.30E-120,\,6.96E-120] & 3.82E-121 & ??? & 17257.61 & ---\\
$\B_{15}$ & 196 & 225 & ??? & 413804 &  5.70E-145 & [5.07E-145,\,6.52E-145] & 1.55E-145 & ??? & 31812.67 & ---\\
\end{tabular*}
\caption{ Overall results; 
$\epsilon=1$, ``swap" indicates it ran out of memory and started
swapping. ``???" indicates that the exact volume is unknown; ``---" indicates it didn't terminate after at least 10h. {\tt VINCI} is used for exact volume computation except Birkhoff polytopes where {\tt birkhoff} is used instead.\label{table:vol_results}}
\end{table}


\begin{table}[t]\centering\tiny
\begin{tabular*}{\linewidth}{@{\extracolsep{\fill}}rr@{~~}r@{~~}rrrr@{}rrr@{~}r@{~}r}
& & & \multicolumn{4}{c}{RDHR} & & \multicolumn{4}{c}{CDHR}\\\cline{4-7}\cline{9-12}\\[-.5em]
$P$ & $d$ & $\epsilon$ & $\mu$ & [min, max] & $(\vol(P)-\mu)$ & VolEsti & & $\mu$ & [min, max] & $(\vol(P)-\mu)$ & VolEsti  \\ 
 &  &  &  &  &   $/\vol(P)$ & (sec) &  &  &  &  $/\vol(P)$ & (sec) \\\hline\\[-.5em]
$\B_5$ & 16 & 1 & 2.27E-07 & [1.66E-07,2.85E-07] &  0.0072 & 22.90 & &  2.25E-07 & [1.87E-07,2.80E-07] & 0.0003 & 4.06\\
$\B_6$ & 25 & 1 & 8.53E-13 & [3.72E-13,1.22E-12] &  0.0982 & 105.96 & &  9.53E-13 & [7.30E-13,1.15E-12] & 0.0083 & 17.26\\
$\B_7$ & 36 & 1 & 2.75E-20 & [1.78E-21,6.71E-20] &  0.4259 & 479.40 & &  4.82E-20 & [3.86E-20,6.18E-20] & 0.0056 & 56.64\\
cube-10 & 10 & 1 & 1022.8 & [944.3951,1103.968] & 0.0012 & 2.03 & &  1026.83 & [970.3117,1096.469] & 0.0027 & 0.34\\
cube-10 & 10 & 0.4 & --  & -- & --  & -- & &  1022.88 & [993.0782,1060.409] & 0.0011 & 2.02\\
cube-20 & 20 & 1 & 1.04E+6 & [9.38E+5,1.14E+6] & 0.0033 & 25.44 & &  1.04E+6 & [9.74E+5,1.12E+6] & 0.0028 & 4.62\\
\end{tabular*} 
\caption{ Experiments with CDHR vs RDHR; $W=10$. 
\label{table:hnr_compare}}
\end{table}

\begin{table}[t]\centering\tiny
\begin{tabular*}{\linewidth}{@{\extracolsep{\fill}}r@{\quad}r@{\quad}rrrrr@{\quad}r@{\quad}|r@{\quad}r@{\quad}|r@{\quad}r@{\quad}r@{\quad}r}
$d$ & $m$ &  \vol$(P)$ & N & $\mu$ & [min, max] & std-dev & $(\vol(P)-\mu)$ & VolEsti  & mem. & VolEsti* & mem. \\
 &  & & & & &  & $/\vol(P)$ & (sec)  & MB & (sec) & MB \\
\hline
10 & 1024 &  2.82E-04 &  9210 &  2.82E-04 & [2.67E-04,3.00E-04] & 5.74E-06 & 0.0001 & 1.58 & 35 & 0.51 & 42\\
12 & 4096 &  8.55E-06 &  11927 & 8.54E-06 & [8.04E-06,8.89E-06] & 1.72E-07 & 0.0010 &  12.21 & 35 & 1.62 & 72\\
14 & 16384 & 1.88E-07 &   14778 &  1.88E-07 & [1.80E-07,1.99E-07] & 4.09E-09 & 0.0006 &  237.22 & 36 & 6.49 & 230\\
16 & 65536 &  3.13E-09 &   17744 &  3.13E-09 & [2.97E-09,3.33E-09] & 6.44E-11 & 0.0004 &   1430.93 & 37 & 32.87 & 992\\
18 & 262144 & 4.09E-11 &  20810 &  4.09E-11 & [3.99E-11,4.29E-11] & 7.19E-13 & 0.0013 & 5791.06& 38 & 188.43 &  4781\\
\end{tabular*} 
\caption{ Experiments with NN for boundary oracle on cross-polytopes;
VolEsti$^*$ uses {\tt flann}; $\epsilon=1$. 
\label{table:flann}}
\end{table}

We implement and experimentally test the above algorithms and methods in the 
software package {\tt VolEsti}.
The code currently consists of around 2.5K lines in C++ and is
open-source\footnote{\tt http://sourceforge.net/projects/randgeom}.
It relies on the {\tt CGAL} library~\cite{CGAL} for its $d$-dimensional kernel
to represent objects such as points and vectors,
for its LP solver~\cite{CGAL_LPSolver},
for the approximate minimum ellipsoid~\cite{CgalBoundingVolumes},
and for generating random points in balls.
We use {\tt Eigen}~\cite{eigenlib} for linear algebra.
The memory consumption is dominated by the list of random points
which needs $O(dN)$ space during the entire execution
of the algorithm (Sect.~\ref{sec:vol}). 
Arithmetic uses the {\tt double} data type of C++,
except from the LP solver, which uses the
\emph{GNU Multiple Precision arithmetic library} to avoid double exponent
overflow. 
We experimented with several pseudo-random number generators in
Boost~\cite{Boostrandom} and chose the fastest, namely mersenne twister generator {\tt mt19937}.
All timings are on an Intel Core i5-2400 $3.1$GHz, $6$MB L2 cache,
$8$GB RAM, 64-bit Debian GNU/Linux.

\vspace{.5em}
\textbf{Data.}
The following polytopes are tested (the first 7 are from the {\tt VINCI} 
webpage):\vspace{-.2cm}
\begin{itemize}\itemsep-.3em
\item cube-$d$:  $\{x=(x_1,\dots,x_d)\, |\, x_i\leq1, x_i\geq-1, x_i\in\RR
\text{ for all } i=1,\dots,d \}$,
\item cross-$d$: cross polytope, the dual of cube, i.e.\
$\mbox{conv}(\{-e_i,e_i,\, i=1,\dots,d\})$,
\item rh-$d$-$m$: polytopes constructed by randomly choosing $m$ hyperplanes tangent to the sphere, 
\item rv-$d$-$n$: dual to rh-$d$-$m$, i.e.\ polytopes with $n$ vertices randomly distributed on the sphere,
\item cc-8-$n$: the $8$-dimensional product of two $4$-dimensional cyclic polyhedra with $n$ vertices, 
\item ccp-$n$: complete cut polytopes on $n$ vertices,
\item Fm-$d$:\ one facet of the metric polytope in dimension $d$,
\item $\Delta$-$d$: the $d$-dimensional simplex $\mbox{conv}(\{e_i,\text{ for } i=0,1,\dots,d\})$,
\item $\Delta$-$d$-$d$: product of two simplices, i.e\ $\{(p, p') \in \RR^{2d}\, |\, p \in \Delta\mbox{-}d, p' \in \Delta\mbox{-}d\}$,
\item skinny-cube-$d$:  $\{x=(x_1,\dots,x_d) \, |\, x_1\leq100, x_1\geq-100, x_i\leq1, x_i\geq-1, x_i\in\RR \, i=2,\dots,d \}$, 
rotated by $30^o$ in the plane defined by the first two coordinate axes,
\item $\B_n$: the $n$-Birkhoff polytope (defined below).
\end{itemize}

Each experiment is repeated 100 times with $\epsilon=1$ unless otherwise stated. 
The reported timing for each experiment is the mean of 100 timings. 
We keep track of and report the \Min\ and the \Max\ computed values,
the mean $\mu$, and the standard deviation.  We measure the accuracy of our method by (\vol$(P)-\mu$)/\vol$(P)$ and (\Max$-$\Min)/$\mu$; unless otherwise stated mean error of approximation refers to the first quantity. 
The reader should not confuse these quantities which refer to the approximation error that computed {\it in practice} with $\epsilon$ which refers to the {\it objective} approximation error.
Comparing the practical and objective approximation error, our method is in practice more accurate than indicated by the theoretical bounds.
In particular, in all experiments all computed values are contained in the
interval $((1-\epsilon)\vol(P),(1+\epsilon)\vol(P))$,
while theoretical results in~\cite{KannanLS97} guarantee only $75\%$ of them.
Actually, the above interval is larger than [\Min, \Max].
In general our experimental results show that our software can approximate
the volume of general polytopes up to dimension 100 in less than 2~hours
with mean approximation error at most 2\% (cf.\ Table~\ref{table:vol_results}).

\vspace{.5em}
\textbf{Random walks and oracles.}
First, we compare the implementations of boundary oracles using membership oracles versus using facet intersection.  
By performing experiments with RDHR our algorithm
approximate the volume of a 10-cube in 42.58~sec using the former,
whereas it runs in~2.03~sec using the latter. 

We compare RDHR to CDHR. The latter take advantage of more efficient boundary oracle implementations as described in Sect.~\ref{sec:OraclSampl}.
Table~\ref{table:hnr_compare} shows that our algorithm using CDHR
becomes faster and more accurate than using RDHR by means of smaller
[\Min,\Max] interval.
Additionally, since CDHR is faster we can increase the accuracy
(decrease $\epsilon$) and obtain even more accurate results than RDHR,
including smaller error $(\vol(P)-\mu)/\vol(P)$. 

Finally, we evaluate our implementations of boundary oracles using {\it duality} and {\it NN search} (Sect.~\ref{subsec:duality}). The motivation comes from the fact that the boundary oracle becomes slow when the number of facets is large, e.g., for cross-$d$, $m=2^d$. 
We consider state-of-the-art NN software:
{\tt CGAL}'s dD Spatial Searching implements kd-trees \cite{CGALKDtree},
{\tt ANN}~\cite{ANN97} implements kd- and BBD-trees,
{\tt LSH} implements Locality Sensitive Hashing~\cite{Andoni08},
and {\tt FLANN}~\cite{Muj11} implements randomized kd-trees.   
We compare them against our oracle running in $O(m)$, on
cross-$17$, $\B_{10}$ and cpp-$7$.
We build two kd-trees per coordinate, i.e.\ one per direction, each
tree storing the dual of the corresponding lower and upper hulls.

Consider point queries.  
{\tt FLANN}, is very fast in high dimensions
(typically $>100$), but lacks theoretical guarantees.
It turns out that {\tt KDTreeSingleIndexParams} on cross-$d$ returns
exact results for all $\varepsilon$ and $d$ tested, since the tree stores vertices of a cube.
Compared to the $O(m)$ oracle, for $\varepsilon=0$ it is 10x slower,
for $\varepsilon=2$ it is competitive, and for $\varepsilon=5$ it lets us approximate
vol(cross-18) with a 40x speed-up, but with extra memory usage
(Table~\ref{table:flann}).   
On other datasets, {\tt FLANN} does not always compute the exact NN 
even for $\varepsilon=0$.
{\tt ANN}, is very fast up to dimension $20$ and
offers theoretical guarantees. For $\varepsilon=0$, it guarantees the exact NN,
but is $>10^3$x slower than our $O(m)$ oracle, though it becomes
significantly faster for $\varepsilon>1$. 
In~\cite{Muja09}, {\tt LSH} is reported to be 10x slower than {\tt FLANN}
and competitive with {\tt ANN}, thus we do test it here.

{\tt CGAL} for point queries is slower than {\tt ANN}, but can be
parametrized to handle hyperplane queries with theoretical guarantees.
Given hyperplane $H$, we set as query point the projection of the origin
on $H$ and as distance-function the inner product between points. 
With the {\tt Sliding\_midpoint} rule and $\varepsilon=0$, this is a bit
(while {\tt ANN} is $1000$x) slower than our boundary oracle for cross-17.
It is important to design methods for which
$\varepsilon>0$ accelerates computation so as to use them with
approximate boundary oracles. 

The above study provides motivation for the design of algorithms that can use approximate boundary queries and hence take advantage of NN software to handle more general polytopes with large number of facets. 
Of particular relevance is the development of efficient methods and
data-structures for approximate hyperplane queries.


\begin{figure*}[t]
	\centering
	\includegraphics[width=.6\textwidth]{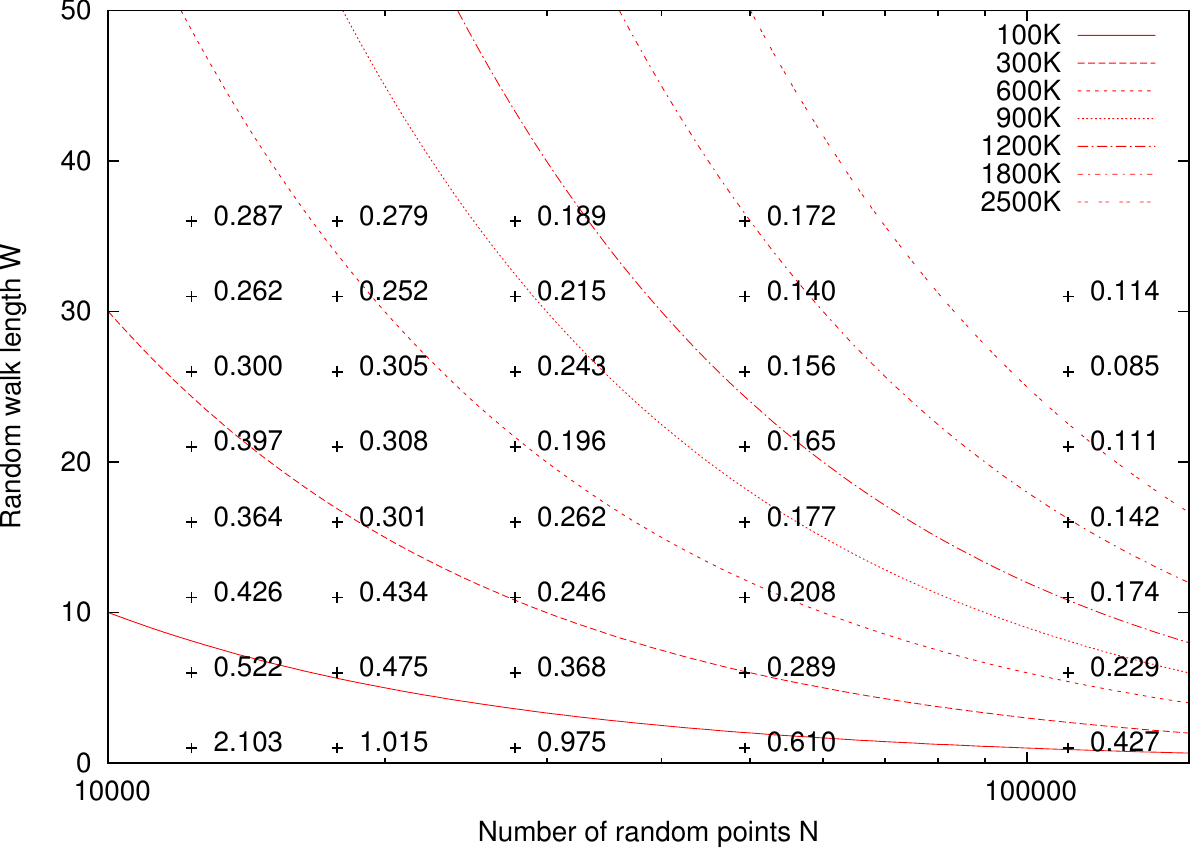}
	\caption{Runtime of {\tt VolEsti}
w.r.t.\ dimension; $\epsilon=1$, y-axis in logscale; 
fitting on cube-$d$ results.}\label{fig:wN}
\end{figure*}

\begin{table}[t!]\centering\scriptsize
\begin{tabular*}{\linewidth}{@{\extracolsep{\fill}}rrrrrrrrrrr}
$P$ & $d$ & $m$ & $W$ & $\mu$ & [\Min,\Max] & std-dev & (\vol$(P)$-$\mu$) & 
(min, max) \\
&  &  & &  & &  & /\vol$(P)$ & /$\mu$ \\
\hline
(*) cube-10 & 10 & 20 & {\bf 10} & 1026.953 & [925.296,1147.101] & 33.91331 & 0.0029 & 0.2160\\
cube-10 & 10 & 20 & 15 & 1024.157 & [928.667,1131.928] & 31.34121 & 0.0002 & 0.1985\\
cube-10 & 10 & 20 & 20 & 1026.910 & [932.118,1144.601] & 30.97023 & 0.0028 & 0.2069\\
 &  &  &  &  &  &  &  & \\
cube-50 & 50 & 100 & 10 & 1.123E+15 & [1.019E+15,1.257E+15] & 4.135E+13 & 0.0022 & 0.2125\\
(*) cube-50 & 50 & 100 & {\bf 15} & 1.131E+15 & [1.039E+15,1.237E+15] & 3.882E+13 & 0.0044 & 0.1744\\
cube-50 & 50 & 100 & 20 & 1.127E+15 & [1.033E+15,1.216E+15] & 3.893E+13 & 0.0007 & 0.1629\\
 &  &  &  &  &  &  &  & \\
cube-100 & 100 & 200 & 10 & 1.278E+30 & [1.165E+30,1.402E+30] & 4.819E+28 & 0.0081 & 0.1856\\
cube-100 & 100 & 200 & 15 & 1.250E+30 & [1.243E+30,1.253E+30] & 4.075E+27 & 0.0140 & 0.0083\\
(*) cube-100 & 100 & 200 & {\bf 20} & 1.263E+30 & [1.190E+30,1.321E+30] & 3.987E+28 & 0.0038 & 0.1038\\
 &  &  &  &  &  &  &  & \\
$\Delta$-20-20 & 40 & 42 & 10 & 1.699E-37 & [1.527E-37,1.881E-37] & 7.670E-39 & 0.0056 & 0.2083\\
(*) $\Delta$-20-20 & 40 & 42 & {\bf 14} & 1.694E-37 & [1.526E-37,1.892E-37] & 7.096E-39 & 0.0025 & 0.2166\\
$\Delta$-20-20 & 40 & 42 & 20 & 1.694E-37 & [1.433E-37,1.836E-37] & 7.006E-39 & 0.0024 & 0.2382\\
 &  &  &  &  &  &  &  & \\
$\Delta$-50-50 & 100 & 102 & 10 & 1.098E-129 & [1.012E-129,1.189E-129] & 4.652E-131 & 0.0154 & 0.1612\\
$\Delta$-50-50 & 100 & 102 & 15 & 1.111E-129 & [1.090E-129,1.139E-129] & 1.610E-131 & 0.0281 & 0.0437\\
(*) $\Delta$-50-50 & 100 & 102 & {\bf 20} & 1.079E-129 & [1.011E-129,1.148E-129] & 3.685E-131 & 0.0015 & 0.1266\\
 &  &  &  &  &  &  &  & \\
$\B_{10}$ & 81 & 100 & 10 & 7.951E-55 & [6.291E-55,9.077E-55] & 8.533E-56 & 0.0946 & 0.3504\\
$\B_{10}$ & 81 & 100 & 15 & 8.124E-55 & [7.451E-55,8.774E-55] & 5.015E-56 & 0.0750 & 0.1629\\
(*) $\B_{10}$ & 81 & 100 & {\bf 20} & 7.489E-55 & [7.398E-55,7.552E-55] & 6.615E-57 & 0.1472 & 0.0106\\
\end{tabular*} 
\caption{ Experiments with varying $W$; $\epsilon=1$.  (*) 
indicate minimum $W$ where either (\vol$(P)$-$\mu$)/\vol$(P)$ or
(min, max)/$\mu$ is $<1\%$.  \label{table:walk_len}}
\end{table}

\begin{table}[t]\centering\scriptsize
\begin{tabular*}{\linewidth}{@{\extracolsep{\fill}}rrrrrrr}
$P$ & \vol$(P)$ & $N$ & $\mu$ & [\Min,\Max] & $\frac{\vol(P)-\mu}{\vol(P)}$ & {\tt VolEsti}(sec) 
\\\hline
rv-8-11 & 3.047E+18 & 6654 & 1.595E+18 & [6.038E+17,3.467E+18] & 0.4766 & 1.48\\
rv-8-11 & 3.047E+18 & 665421 & 3.134E+18 & [3.134E+18,3.134E+18] & 0.0283 & 157.46\\
(*) rv-8-11 & 3.047E+18 & 6654 & 3.052E+18 & [2.755E+18,3.383E+18] & 0.0013 & 1.34\\
skinny-cube-10 & 1.024E+05 & 9210 & 5.175E+04 & [2.147E+04,1.228E+05] & 0.4946 & 0.69\\
(*) skinny-cube-10 & 1.024E+05 & 9210 & 1.029E+05 & [8.445E+04,1.149E+05] & 0.0050 & 0.71\\
skinny-cube-20 & 1.049E+08 & 23965 & 4.193E+07 & [2.497E+07,7.259E+07] & 0.6001 & 5.59\\
(*) skinny-cube-20 & 1.049E+08 & 23965 & 1.040E+08 & [8.458E+07,1.163E+08] & 0.0084 & 6.70\\
\end{tabular*} 
\caption{ Experiments with rounding; (*): means that we use rounding.
\label{table:rounding}}
\end{table} 

\textbf{Choice of parameters and rounding.}
We consider two crucial parameters,  
the length of a random walk, denoted by $W$, and approximation
$\epsilon$, which determines the number $N$ of random points. 
We set $W=\lfloor 10+d/10\rfloor$.
Our experiments indicate that, with this choice, either
(\vol$(P)$-$\mu$)/\vol$(P)$ or $(min, max)/\mu$ is $<1\%$
up to $d= 100$~(Table~\ref{table:walk_len}). 
Moreover, for higher $W$ the improvement in accuracy is not significant,
which supports the claim that asymptotic bounds are unrealistically high.
Fig.~\ref{fig:wN} correlates runtime (expressed by $NW$) and
accuracy (expressed by $(min, max)/\mu$ which actually measures 
some ``deviation") to $W$ and $\epsilon$ (expressed by $N$).
A positive observation is that accuracy tightly correlates with runtime:
e.g., accuracy values close to or beyond~1 lie under the curve $NW=10^5$,
and those rounded to $\le 0.3$ lie roughly above $NW=3\cdot 10^5$.
It also shows that,
increasing $W$ converges faster than increasing $N$ to a value beyond which the improvement in accuracy is not significant.

To experimentally test the effect of {\it rounding} we construct skinny
hypercubes skinny-cube-d.  
We rotate them to avoid CDHR taking unfair advantage of the degenerate
situation where the long edge is parallel to an axis. 
Table~\ref{table:rounding} on these and other polytopes shows that
rounding reduces approximation error by~2 orders of magnitude.
Without rounding, for polytope rv-8-11
one needs to multiply $N$ (thus runtime) by 100 in order to achieve 
approximation error same as with rounding.  


\begin{table}[t]\centering\footnotesize
\begin{tabular}{r|rrrr|rrrr|rrrr}
\multirow{2}{.5cm}{$P$:} & \multicolumn{4}{c}{rv-15-} & \multicolumn{4}{c}{rv-10-} & \multicolumn{4}{c}{cube-} \\
& 30 & 40 & 50 & 60 & 100 & 150 & 200 & 250 & 7 & 8 & 9 & 10 \\\hline
time (sec) & 7.7 & 82.8 & 473.3 & swap & 37.3 & 107.8 & 282.5 & 449.0 & 0.1 & 2.2 & 119.5 & $>5$h\\
\end{tabular} 
\caption{ Experiments with {\tt qhull}; ``swap" indicates it
ran out of memory and started swapping; ``$>$5h" indicates it did not
terminate after 5 hours.  \label{table:qhull_vol}}
\end{table}

\begin{table}[t!]\centering\tiny
\begin{tabular*}{\linewidth}{@{\extracolsep{\fill}}r@{\,}|@{\,}r@{\,}
r@{\,}r@{\,}r@{\,}r@{\,}|@{\,}r@{\,}r@{\,}r@{\,}r@{\,}r}
& \multicolumn{5}{c}{software of~\cite{CousinsV13}} & \multicolumn{5}{c}{{\tt Volesti}}\\
 $P$  & [min, max]  & std-dev & {\tiny \multirow{2}{*}{$\frac{\vol(P)-\mu}{\vol(P)}$}} & \# total & time(sec)
  & [min, max]  & std-dev & \multirow{2}{*}{$\frac{\vol(P)-\mu}{\vol(P)}$} & \# total & time(sec)\\
 &&&&steps&&&&&steps& \\\hline
cube-20  & [5.11E+05,\,1.55E+06] & 1.67E+05 & 0.0198 & 7.96E+04 & 21.48 & [9.74E+05,\,1.12E+06] & 3.15E+04 & 0.0028 & 3.61E+06 &  4.62\\
cube-30  & [6.75E+08,\,1.45E+09] & 1.72E+08 & 0.0440 & 2.22E+05 &  49.24 & [9.91E+08,\,1.16E+09] & 3.89E+07 & 0.0039 & 1.21E+07 &  17.96\\
cube-40  & [7.90E+11,\,1.38E+12] & 1.67E+11 & 0.0731 & 4.30E+05 &  88.09 & [1.01E+12,\,1.23E+12] & 4.46E+10 & 0.0039 & 2.84E+07 &  50.72\\
cube-50  & [8.75E+14,\,1.45E+15] & 1.43E+14 & 0.0327 & 7.16E+05 &  148.06 & [1.00E+15,\,1.25E+15] & 4.39E+13 & 0.0007 & 5.49E+07 &  117.51\\
cube-60  & [8.89E+17,\,1.43E+18] & 1.64E+17 & 0.0473 & 1.15E+06 &  229.33 & [1.06E+18,\,1.27E+18] & 4.00E+16 & 0.0051 & 9.42E+07 &  222.10\\
cube-70  & [9.01E+20,\,1.36E+21] & 1.49E+20 & 0.0707 & 1.66E+06 &  427.82 & [1.02E+21,\,1.32E+21] & 5.42E+19 & 0.0013 & 1.49E+08 &  358.93\\
cube-80  & [9.30E+23,\,1.36E+24] & 1.46E+23 & 0.1145 & 2.30E+06 &  531.46 & [1.13E+24,\,1.30E+24] & 4.42E+22 & 0.0009 & 2.21E+08 &  582.19\\
cube-90  & [1.07E+27,\,1.88E+27] & 2.20E+26 & 0.0394 & 3.30E+06 &  701.54 & [1.09E+27,\,1.44E+27] & 5.18E+25 & 0.0019 & 3.15E+08 &  875.69\\
cube-100 & [9.53E+29,\,1.64E+30] & 1.93E+29 & 0.0357 & 4.19E+06 &  884.43 & [1.17E+30,\,1.40E+30] & 4.82E+28 & 0.0081 & 4.33E+08 &  1285.08\\
$\B_{8}$   & [2.12E-23,\,2.45E-22] & 6.25E-23 & 0.3970 & 9.31E+05 &  221.30 & [4.05E-23,\,7.32E-24] & 1.93E+04 & 0.0092 & 1.01E+08 & 192.97\\
$\B_{9}$   & [1.54E-33,\,2.77E-33] & 3.71E-34 & 0.1830 & 2.05E+06 &  420.07 & [2.23E-33,\,3.07E-33] & 2.13E-34 & 0.0069 & 2.27E+08 & 499.56\\
$\B_{10}$  & [3.39E-46,\,1.92E-45] & 4.75E-46 & 0.1207 & 3.69E+06 &  691.97 & [7.97E-46,\,9.96E-46] & 4.99E-47 & 0.0152 & 4.62E+08 & 1034.74\\
\end{tabular*} 
\caption{ Comparison of the software~\cite{CousinsV13} vs {\tt VolEsti}; each experiment is run 10 times, total steps refer to the mean of the total number of Hit-and-run steps in each execution.\label{table:matlab}}
\end{table} 

\vspace{.5em}
\textbf{Other software.}
Exact volume computation concerns software computing the exact value of
the volume, up to round-off errors in case it uses floating point arithmetic. 
We mainly test against {\tt VINCI}\,1.0.5~\cite{vinci}, which implements
state-of-the art algorithms, cf.\ Table~\ref{table:vol_results}.
For H-polytopes, the method based on Lawrence's general formula 
is numerically unstable resulting in wrong results in many examples~\cite{FukVolume00},
and thus was excluded.
Therefore, we focused on Lasserre's method. 
For all polytopes there is a threshold dimension for which
{\tt VINCI} cannot compute the volume:
it takes a lot of time (e.g.\ $>$ 4 hrs for cube-20) and 
consumes all system memory, thus starts swapping.

{\tt LRS} 
is not useful for H-polytopes as stated on its webpage:
``If the volume option
is applied to an H-representation, the results are not predictable."
{\tt Latte} implements the same decomposition methods
as {\tt VINCI}; it is less prone to round-off error but slower~\cite{Latte}.
{\tt Normaliz} 
applies triangulation:
it handles cubes for $d\le 10$, in $<1$~min,
but for $d=15$, it did not terminate after 5~hours.
{\tt Qhull} handles V-polytopes
but does not terminate for cube-10 nor random polytope rv-15-60
(Table~\ref{table:qhull_vol}). 
This should be juxtaposed to the duals, namely
our software approximates the volume of cross-10 in 2 sec with $<1\%$ error and 
rh-15-60 in 3.44 sec.
A general conclusion for exact software is that it cannot handle
$d> 15$.

We compare with the most relevant approximation method, namely the
{\tt Matlab} implementation of~\cite{CousinsV13_matlab} for bodies
represented as the intersection of an H-polytope and an ellipsoid.
They report that the code is optimized to achieve about $75\%$ success
rate for bodies of dimension $\le 100$ and $\epsilon\in [0.1, 0.2]$ (not to be confused with the $\epsilon$ of our method).
Testing~\cite{CousinsV13_matlab} with default options and
$\epsilon=0.1$, our implementation with $\epsilon=1$ runs faster for $d<80$, performs  roughly $100$ times more total Hit-and-run steps and returns significantly
more accurate results, e.g.\ from $4$ to $100$ times smaller error
on cube-$d$ when $d>70$, and from $5$ to $80$ times on Birkhoff polytopes (Table~\ref{table:matlab}).

\begin{table}[t!]\centering\small
\begin{tabular}{rr|rrrrrrrr}
& n & 3 & 4 & 5 & 6 & 7 & 8 & 9 & 10\\\hline
\multirow{2}{1cm}{$\frac{\text{estimate}}{\text{actual}}$} & \cite{GMcKay09} & 1.25408 & 1.22556 & 1.19608 & 1.17258 & 1.15403 & 1.13910 & 1.12684 & 1.11627\\
& VolEsti & 0.99485 & 1.09315 & 1.00029 & 1.00830 & 1.00564 & 0.99440 & 0.99313 & 1.01525\\  
\end{tabular} 
\caption{ Comparison between asymptotic and experimental approximation of the volume of $\B_n$.
\label{table:birk_asymptotic}}
\end{table}

\vspace{.5em}
\textbf{Birkhoff polytopes}
are well studied in combinatorial geometry and offer an important benchmark.
The $n$-th Birkhoff polytope 
$\B_n =$ $\{ x \in \RR^{n\times n}\; |\, \ x_{ij} \geq 0,\;
\sum_i x_{ij} = 1,\; \sum_j x_{ij} = 1,\; 1 \leq i \leq n \},$
also described as the polytope of the 
perfect matchings of the complete bipartite graph $K_{n,n}$, 
the polytope of the $n\times n$ doubly stochastic matrices, 
and the Newton polytope of the determinant.
In~\cite{BeckPixton03}, they present a complex-analytic method for 
this volume, implemented in package {\tt birkhoff}, which has managed
to compute \vol$(\B_{10})$ in parallel execution, 
which corresponds to a single processor running at 1~GHz for almost 17~years.

First, $\dim\B_n = {n^2-2n+1}$: we project $\B_n$ to
a subspace of this dimension.
Our software, with $\epsilon=1$,  computes the volume of polytopes up to $\B_{10}$
in $<1$~hour with mean error of $\le 2\%$ (Table~\ref{table:vol_results}).
The computed approximation values improves upon the best known upper bounds on $\vol(\B_n)$, obtained through the asymptotic formula of~\cite{GMcKay09}, cf.\ Table~\ref{table:birk_asymptotic}.
By setting $\epsilon=.5$ we obtain an error of $0.7\%$ for \vol$(\B_{10})$,
in $6$~hours. The computed approximation of the volume has two correct digits, i.e.\ its first two digits equal to the ones of the exact volume.
More interestingly, using $\epsilon=1$ we compute, in $<9$ hours, an approximation as well as an interval of values for \vol$(\B_{11})$,\dots, \vol$(\B_{15})$,
whose exact values are unknown (Table~\ref{table:vol_results}).

\section{Further work}\label{sec:further}

NN search seems promising and could accelerate our
code, especially if it were performed approximately with
hyperplane queries.
Producing (almost) uniform point samples
is of independent interest in machine learning, including
sampling contingency tables and learning the p-value. We plan
to exploit such applications of our software. 
We may also study sampling for special polytopes such as Birkhoff.
It is straightforward to parallelize certain aspects of the algorithm, 
such as random walks assigning each thread to a processor,
though other aspects, such as the algorithm's phases,
require more sophisticated parallelization.
Our original motivation and ultimate goal is to extend these methods to
V-polytopes represented by an optimization oracle.

\section{Acknowledgments}
This work is  
co-financed by the European Union (European Social Fund - ESF) and Greek national funds through the Operational Program ``Education and Lifelong Learning" of the National Strategic Reference Framework (NSRF) - Research Funding Program: THALIS - UOA (MIS 375891).
The authors acknowledge discussions with Matthias Beck on Birkhoff polytopes,
and Andreas Enge on {\tt VINCI}, and help with experiments
by Ioannis Psarros and Georgios Samaras, students at UoA.

\bibliographystyle{plain} 
\bibliography{../bibliography} 

\end{document}